\documentclass[11pt,a4paper]{article}

\usepackage{amsfonts,amssymb,amsthm}
\usepackage{cite} 
\usepackage[T2A]{fontenc}
\usepackage[cp1251]{inputenc}
\usepackage[english,russian]{babel}
\usepackage{amsmath}%
\usepackage[unicode]{hyperref}
\usepackage[left=1.5cm,right=1.5cm,
    top=2cm,bottom=3cm,bindingoffset=0cm]{geometry}

\allowdisplaybreaks
\numberwithin{equation}{section}

\title{\bf Symmetries of Toda type 3D lattices\footnote{International Conference <<Integrable systems and their applications>>, Sochi--2024}}
\author{\bf I.T.Habibullin, A.R.Khakimova} 

\begin{document}
\maketitle

\begin{center}
Institute of Mathematics, Ufa Scientific Center, Russian Academy of Sciences\\
 
\end{center}

\abstract {The duality between a class of the Davey-Stewartson type coupled systems and a class of two-dimensional Toda type lattices is discussed. For the recently found integrable lattice the hierarchy of symmetries is described. Second and third order symmetries are presented in explicit form. Corresponding coupled systems are given. An original method for constructing exact solutions to coupled systems is suggested based on the Darboux integrable reductions of the dressing chains. Some new solutions for coupled systems related to the Volterra lattice are presented as illustrative examples.} 

\vspace{0.5cm}


\vspace{0.5cm}

\section{Introduction}

It was noted in paper \cite{ShabatYamilov} that there is a duality between two-dimensional lattices and coupled systems of the Davey-Stewartson type. Namely, coupled systems are generalized symmetries for lattices. Lattices in turn provide dressing chains for coupled systems. Using this duality, Shabat and Yamilov found a class of coupled systems corresponding to a known list of lattices containing six models. Among them, they introduced such an important integrable equation as a spatially two-dimensional generalization of the Heisenberg equation. The results of this work are important both from the point of view of the integrable classification of both types of equations and from the point of view of finding their explicit particular solutions. However, the scientific direction initiated in \cite{ShabatYamilov} did not find effective application for a long time due to problems with non-local variables arising in the theory of multidimensional integrable systems (see, or instance, \cite{LeznovShabatYamilov}). In our recent papers \cite{Habibullin13}, \cite{HabibullinPoptsova18}   it was discovered that integrable discrete and differential-difference equations with three independent variables admit infinite hierarchies of reductions in the form of Darboux-integrable systems of hyperbolic equations of dimension 1+1. This circumstance allowed us to overcome the problem of nonlocal variables in a number of cases (see  \cite{HabibullinKuznetsova20}, \cite{Kuznetsova19}, \cite{HabibullinKhakimova21}).

Let us consider illustrative examples. It was shown in \cite{ShabatYamilov} that the two-dimensional Volterra chain
\begin{equation}\label{Volterra}
u_{ny}=u_n(v_n-v_{n+1}), \qquad v_{nx}=v_n(u_{n-1}-u_n)
\end{equation}
has a symmetry (coupled system) of the following form
\begin{equation}\label{csVolterra}
\begin{aligned}
&u_{nt}=u_{nxx}+\left(u_n^2+2u_nV_n\right)_x,\\
&v_{nt}=-v_{nxx}+\left(V^2_n\right)_y+(2u_nv_n)_x, \quad V_{ny}=v_{nx}.
\end{aligned}
\end{equation}
It is proven in \cite{ShabatYamilov} that the lattice provides an invertible B\"acklund transformation for  system \eqref{csVolterra}
\begin{equation*}%\label{Backlund1}
v_{n-1}=v_{n}-(\ln u_{n})_y,\, u_{n-1}=u_{n}+(\ln v_{n-1})_x, \, V_{n-1}=V_{n}-(\ln u_{n})_y.
\end{equation*}

The Volterra chain admits a large class of symmetries. For instance, one can easily derive another coupled system of the second order from \eqref{csVolterra} by using the involutions $x\leftrightarrow -y$, $t\leftrightarrow \tau$, $u\leftrightarrow v$, $ U\leftrightarrow  V$, $n \leftrightarrow -n$
\begin{equation}%\label{csVolterra2}
\begin{aligned}
&u_{n\tau}=u_{nyy}+\left( U^2_n\right)_x +(2u_nv_n)_y,\\
&v_{n\tau}=-v_{nyy}+\left(v^2_n+2v_n U_n\right)_y, \quad  U_{nx}=u_{ny}.
\end{aligned}
\end{equation}
\begin{equation*}%\label{Backlund1}
u_{n+1}=u_{n}+(\ln v_{n})_x, \, v_{n+1}=v_{n}+(\ln u_{n+1})_y,\, U_{n+1}=U_{n}+(\ln v_{n})_y
\end{equation*}
By taking linear combinations of two symmetries given above we find a more complicated symmetry
\begin{equation}\label{csVolterra3}
\begin{aligned}
&u_{s}=\lambda u_{xx}+\mu u_{yy}+\lambda\left(u^2+2uV\right)_x    +\mu\left( U^2\right)_x +\mu(2uv)_y,\\
&\qquad V_y=v_x,\quad  \lambda\neq0\\
&v_{s}=\ -\lambda v_{xx} -\mu v_{yy}+\lambda \left(V^2\right)_y+\lambda(2uv)_x +\mu\left(v^2+2v U\right)_y,\\
& \qquad  U_x=u_y, \quad\mu\neq0.
\end{aligned}
\end{equation}

In view of the duality between these two classes, it can be concluded that the presence of a complete list of integrable equations of one of the classes of models would allow obtaining a complete list of integrable representatives of the other class. However, to date, the classification problem has not been solved for either of these two classes.

%\vspace{0.5cm}

We have recently made some progress in the problem of classifying lattices. An algorithm for integrable classification of two-dimensional lattices has been proposed, based on the concept of Darboux-integrable finite-field reductions.

The problem of description of the integrable equations of the form
\begin{equation}  \label{todatype}
u_{n,xy} = f(u_{n+1},u_{n},u_{n-1}, u_{n,x},u_{n,y})
\end{equation}
was reduced to the problem of describing all functions $f$ such that hyperbolic systems
\begin{equation}\label{reduc}
\begin{aligned}
&u_{1,xy}=f_1\left(u_1,u_2,u_{1,x},u_{1,y}\right),\\
&u_{j,xy}=f\left(u_{j+1},u_j,u_{j-1},u_{j,x},u_{j,y}\right), \quad 1<j<m, \\
&u_{m,xy}=f_2\left(u_m,u_{m-1},u_{m,x},u_{m,y}\right)
\end{aligned}
\end{equation}
are integrable in the sense of Darboux for arbitrary integer $m\geq2$ for a suitable choice of the functions $f_1=f_1\left(u_1,u_2,u_{1,x},u_{1,y}\right)$ and $f_2=f_2\left(u_m,u_{m-1},u_{m,x},u_{m,y}\right)$. {\bf The problem of complete classification in general remains open.} But it is solved in the quasilinear case.

For lattices having the following particular quasilinear form
\begin{equation}\label{0}
u_{n,xy}=A_1u_{n,x}u_{n,y}+A_2u_{n,x}+A_3u_{n,y}+A_4    
\end{equation}
where the coefficients depend on the dynamical variables $A_i=A_i(u_{n+1},u_n,u_{n-1})$ for $i=1,2,3,4$, the problem is solved in \cite{HabibullinPoptsova18}, \cite{HabibullinKuznetsova20}, \cite{Kuznetsova19}. Here
we present a list of lattices of class \eqref{0} that passed the test formulated above. This list
is complete up to point transformations.

\begin{itemize}
\item[$(E1)$] $u_{n,xy} = e^{u_{n+1} - 2 u_n + u_{n-1} },$
\item[$(E2)$] $u_{n,xy} = e^{u_{n+1}} - 2 e^{u_n} + e^{u_{n-1}},$
\item[$(E3)$] $u_{n,xy} = e^{u_{n+1}-{u_n}} -  e^{u_n-u_{n-1}},$
\item[$(E4)$] $u_{n,xy} = \left(u_{n+1} - 2 u_n + u_{n-1}  \right) u_{n,x}, $
\item[$(E5)$] $u_{n,xy} = \left(e^{u_{n+1}-{u_n}} -  e^{u_n-u_{n-1}}\right)u_{n,x},$
\item[$(E6)$] $u_{n,xy}=\alpha_nu_{n,x}u_{n,y}, \quad \alpha_n = \frac{1}{u_n - u_{n-1}} - \frac{1}{u_{n+1}-u_n}$
\item[$(E7)$] $u_{n,xy} = \alpha_n(u_{n,x} + u^2_n - 1)(u_{n,y} + u^2_n - 1) - 2 u_n(u_{n,x}+u_{n,y}+u^2_n - 1).$
\end{itemize}

The list of the coupled systems of the DS type corresponding to the lattices $E1$-$E6$ is given in \cite{ShabatYamilov}.

Below we will focus on the following three points:

1) Finding the symmetries for the novel lattice $E7$
\begin{eqnarray}
u_{n,xy} = \alpha_n(u_{n,x} + u^2_n - 1)(u_{n,y} + u^2_n - 1) - 2 u_n(u_{n,x}+u_{n,y}+u^2_n - 1), \label{newlattice}\\
\alpha_n = \frac{1}{u_n - u_{n-1}} - \frac{1}{u_{n+1}-u_n}.\nonumber
\end{eqnarray}

2) Derivation of the corresponding DS type coupled system.

3) Discussion on an approach for constructing particular solutions to the DS-type coupled systems via dressing chain. %(This problem was formulated in [A. N. Leznov, A. B. Shabat, R. I. Yamilov, “Canonical transformations generated by shifts in nonlinear lattices”, Phys. Lett. A, 174:5–6 (1993), 397–402]). 

\section{Symmetries of a three-field lattice}

First, we discuss the problem of constructing symmetries for lattices. It is convenient to start with a three-field lattice of the form
\begin{eqnarray}
&&a_{n,y}=a_n(b_n-b_{n+1}+u_n-u_{n+1}),\nonumber\\
&&b_{n,x}=b_n(a_{n-1}-a_n+u_{n}-u_{n-1}), \label{gVolterra} \\
&&u_{n,y}-a_n(u_{n}-u_{n+1})=u_{n,x}+b_n(u_{n}-u_{n-1}) \nonumber 
\end{eqnarray} 
with the sought functions $a_n$, $b_n$ and $u_n$. Evidently lattice \eqref{gVolterra} is reduced to the Volterra chain under constraint $u_n=0$.
%\begin{equation}\label{Volterra}
%a_{ny}=a_n(b_n-b_{n+1}), \qquad b_{nx}=b_n(a_{n-1}-a_n).
%\end{equation}
Therefore, it can be considered as a three-field generalization of the Volterra chain \eqref{Volterra}.
Another reduction admitted by \eqref{gVolterra} is given by the relations
\begin{equation}\label{reduction}
a_n=\frac{u_{n,x}-u^2_n-1}{u_{n+1}-u_n}, \qquad b_n=\frac{u_{n,y}-u^2_n-1}{u_n-u_{n-1}}.
\end{equation}
In this case the lattice is transformed into equation \eqref{newlattice}. 

The following system of the linear equations (cf. \cite{Kuznetsova21})
\begin{equation}\label{Lax7}
\begin{aligned}
&{\psi}_{n,x}=a_n\left(\psi_{n+1}-\psi_n\right)+u_n\psi_n,\\
&{\psi}_{n,y}=b_n\left(\psi_n-\psi_{n-1}\right)+u_n\psi_n.
\end{aligned}
\end{equation}
provide a Lax pair for the lattice \eqref{gVolterra}. In other words this overdetermined system is compatible if only if  the coefficients $a_n$, $b_n$, $u_n$ solve lattice \eqref{gVolterra}. This fact can be reformulated in terms of the operators
\begin{equation}\label{operators}
\partial_x-B_1 \quad \mbox{and} \quad \partial_y-C_1,
\end{equation}
where $B_1=a_nT-a_n+u_n$ and $C_1=b_n+u_n-b_nT^{-1}$. The shift operator $T$ acts due to the rule $Ty(n)=y(n+1)$. Symbols $\partial_x$ and $\partial_y$  stand for the operators of total differentiation with respect to the variables $x$ and $y$ correspondingly. Actually functions $a_n$, $b_n$, $u_n$ satisfy \eqref{gVolterra} iff the operators \eqref{operators}  commute. 

To construct symmetries of \eqref{gVolterra} we use the method based on the concept of Lax pairs (see, for instance,  \cite{ShabatYamilov}, \cite{Ueno},).
First, we need to describe the class of nonlocal variables on which the symmetries depend. For this purpose, we consider equations
\begin{equation}\label{firstequation}
[\partial_x-B_1, L]=0, \qquad [\partial_y-C_1, M]=0,
\end{equation}
where the sought objects $L$ and $M$ are operators represented as formal power series of the shift operator~$T$
\begin{equation}\label{formalseries}
L=\sum^1_{i=-\infty} \alpha^{(i)}_nT^{i},\qquad M=\sum^{+\infty}_{i=-1} \beta^{(i)}_nT^{i},
\end{equation}
It is supposed that the first summands of the series are taken as follows
\begin{equation}\label{firscoefficients}
\alpha^{(1)}_n=-a_n,\qquad \beta^{(-1)}_n=b_n.
\end{equation}
The other coefficients are found from equations obtained by comparing the factors in front of the powers of $T$. Actually here we have to solve some linear equations, that generate nonlocalities. For example, the nonlocal variables $H_n:=\alpha^{(0)}_n$ and $V_n:=-\alpha^{(-1)}_n$ are found from equations
\begin{equation}\label{nonlocalitiesL0}
(1-T)H_n=D_x\log a_nb_{n+1}, \qquad D_yH_n=(1-T)a_{n-1}b_n
\end{equation}
and equations
\begin{equation}\label{nonlocalitiesL-1}
(1-T)V_na_{n-1}=D_xH_n, \qquad D_yV_na_{n-1}=D_xa_{n-1}b_n,
\end{equation} 
 respectively.
The coefficients $Q_n:=\beta^{(0)}_n$ and $U_n:=\beta^{(1)}_n$ of the series $M$ are obtained in a similar way
\begin{equation}\label{nonlocalitiesM0}
D_xQ_n=(T-1)a_{n-1}b_{n}, \qquad (T-1)Q_n=D_y\log a_nb_{n+1}
\end{equation}
and, correspondingly, 
\begin{equation}\label{nonlocalitiesM1}
D_xU_nb_{n+1}=-D_ya_nb_{n+1}, \qquad (1-T)b_{n+1}U_n=D_yQ_{n+1}.
\end{equation} 

Thus coefficients of the formal series $L$ and $M$ generate an infinite sequence of the nonlocal variables. It is easy to verify that any positive power of each series satisfies an equation similar to \eqref{firstequation} 
\begin{equation}\label{firstequation-k}
[\partial_x-B_1, L^k]=0, \qquad [\partial_y-C_1, M^k]=0,
\end{equation}
We define new operators $B_k$ and $C_k$ for $k\geq2$ due to the rule
\begin{equation}\label{+-}
B_k=(L^k)_+, \qquad C_k=(M^k)_-,
\end{equation}
Let us explain the meanings of the symbols in \eqref{+-}. Assume that
\begin{equation}\label{formalseries_k}
L^k=\sum^k_{i=-\infty} \alpha^{(i,k)}_nT^{i},\qquad M^k=\sum^{+\infty}_{i=-k} \beta^{(i,k)}_nT^{i},
\end{equation}
then we have
\begin{equation}\label{Lk+Mk-}
(L^k)_+=\sum^k_{i=1} \alpha^{(i,k)}_nT^{i}-\sum^k_{i=1} \alpha^{(i,k)}_n,\qquad (M^k)_-=\sum^{-1}_{i=-k} \beta^{(i,k)}_nT^{i}-\sum^{-1}_{i=-k} \beta^{(i,k)}_n,
\end{equation}
It is important that transformations $P_{\pm}$ acting as $P_+:L^k\rightarrow (L^k)_+$ and $P_-:M^k\rightarrow (M^k)_-$ define projection operators.
Note that such methods of truncating formal series differ from the standard methods usually used when searching for higher symmetries of chains in 3D (see, for example, \cite{ShabatYamilov}, \cite{Ueno}).

Such a rule for choosing the polynomial parts of the power series is related to the fact that the basic operators can be written in the following form
$$D_x-B_1=D_x-a_n\Delta_+-u_n,\quad D_y-C_1=D_y+b_n\Delta_--u_n,$$
where $\Delta_{\pm}=T^{\pm1}-1$ are operators of the forward/backward discrete derivatives. The projection operators are defined in such a way that the operators $B_k$ and $C_k$ can be represented as follows
$$B_k=\sum^k_{i=1} \bar\alpha^{(i,k)}_n\Delta_+^{i},\quad C_k=\sum^k_{i=1} \bar\beta^{(i,k)}_n\Delta_-^{i}. $$
Therefore, all operators involved are polynomials of the  operators $\Delta_+$ or $\Delta_-$.

Let us take two copies of time flows $x_2, x_3,x_4,\dots,$ $y_2, y_3, y_4, \dots$. Assume that $x_1:=x$, $y_1:=y$. We define a hierarchy of symmetries of the nonlinear system corresponding to these flows by specifying Lax-type representations (cf. \cite{Ueno})
\begin{equation}\label{Lax}
\begin{aligned}
&\partial_{x_k}L=[B_k,L],\qquad   \partial_{x_k}M=[B_k,M],\\
&\partial_{y_k}L=[C_k,L],\qquad   \partial_{y_k}M=[C_k,M].
\end{aligned}
\end{equation}
To search for symmetries, we move, following the method described in \cite{Ueno}, from the Lax-type representation to the Zakharov-Shabat-type equations
\begin{equation}\label{ZakharovShabat}
\begin{aligned}
&\partial_{x_k}B_m-\partial_{x_m}B_k+[B_m,B_k]=0,\\
&\partial_{y_k}C_m-\partial_{y_m}C_k+[C_m,C_k]=0,\\
&\partial_{y_k}B_m-\partial_{x_m}C_k+[B_m,C_k]=0.
\end{aligned}
\end{equation}

{\bf Theorem 1.} Equations of system 
\eqref{ZakharovShabat} are self consistent, i.e. they produce dynamical systems for arbitrary positive integers $k$ and $m$.

\vspace{0.2cm}

Therefore the three-field lattice \eqref{gVolterra} admits an infinite hierarchy of symmetries. The same is true for the lattice $E7$.

\section{Examples of symmetries}

\textbf{3.1.} We use the following two pairs of equations
\begin{equation}\label{ZakharovShabatx2}
\begin{aligned}
&\partial_{x_2}B_1-\partial_{x_1}B_2+[B_1,B_2]=0,\\
&\partial_{x_2}C_1-\partial_{y_1}B_2+[C_1,B_2]=0
\end{aligned}
\end{equation}
and
\begin{equation}\label{ZakharovShabaty2}
\begin{aligned}
&\partial_{y_2}C_1-\partial_{y_1}C_1+[C_1,C_2]=0,\\
&\partial_{y_2}B_1-\partial_{x_1}C_1+[B_1,C_2]=0
\end{aligned}
\end{equation}
to construct the symmetries of \eqref{gVolterra} corresponding to times $x_2$ and $y_2$. 

We search for the operator $B_2$ using the definition \eqref{+-} and the rule \eqref{Lk+Mk-}. As a result of simple calculations we find
\begin{equation}\label{B2}
B_2=a_na_{n+1}T^2-a_n(H_n+H_{n+1})T-a_na_{n+1}+a_n(H_n+H_{n+1}).
\end{equation}

Operator $B_2$ is easily rewritten in terms of the operator $\Delta_+$
\begin{equation}\label{B2delta}
B_2=a_na_{n+1}\Delta_+^2+(2a_na_{n+1}-a_nH_n-a_nH_{n+1})\Delta_++a_na_{n+1}.
\end{equation}

Then we substitute the operators $B_1$, $C_1$ and $B_2$ defined above into system \eqref{ZakharovShabatx2} and after some simplification due to \eqref{gVolterra} and \eqref{nonlocalitiesL0} we arrive at an overdetermined system of nonlinear equations:
\begin{align*}
&u_{n,x_2}-a_{n,x_2}+R_{n,x_1}=0,\\
&\frac{b_{n,x_2}}{b_n}-R_{n-1}+R_{n}=0,\\
&\frac{a_{n,x_2}}{a_n}+\left(\log a_nb_{n+1}\right)_{x_1}\left(H_{n+1}+H_n\right)+H_{n+1,x_1}+H_{n,x_1}+R_{n+1}-R_{n}=0,\\
&b_{n,x_2}+u_{n,x_2}-R_{n,y_1}+a_{n-1}b_{n}\left(H_{n}+H_{n-1}\right)-a_{n}b_{n+1}\left(H_{n+1}+H_{n}\right)=0,
\end{align*}
where according to \eqref{Lk+Mk-} we have $R_n=a_na_{n+1}-a_n(H_n+H_{n+1})$. 

Despite the fact that we have obtained four equations for three sought functions, the system is consistent, since the rule for terminating formal series given above in \eqref{Lk+Mk-} was chosen in a suitable way, guaranteeing the consistency of the dynamical system.

After simple calculations, we obtain an explicit expression for the desired symmetry  of  system \eqref{gVolterra} in the direction of $x_1$:
\begin{equation}\label{symmfor3}
\begin{aligned}
a_{n,x_2}=&a_{n,x_1x_1}+2a_na_{n,x_1}-2\left(a_nH_n\right)_{x_1}-a_n\left(u_{n,x_1}-u_{n+1,x_1}\right)+a_n\left(u_{n+1}-u_{n}\right)^2\\
&+a_{n+1}a_n\left(u_{n+2}-u_{n+1}\right)-\left(u_{n+1}-u_{n}\right)\left(2a_nH_n-a_n^2-2a_{n,x_1}+a_n\right),\\
b_{n,x_2}=&2b_n\left(a_n-a_{n-1}\right)H_n-b_n\left(a_{n,x_1}+a_{n-1,x_1}\right)-b_n\left(a_n-a_{n-1}\right)^2\\
&-a_nb_n(u_{n+1}-u_n)-a_{n-1}b_n(u_{n}-u_{n-1})+b_n(u_n-u_{n-1}),\\
u_{n,x_2}=&u_{n,x_1}-2a_nH_n\left(u_{n+1}-u_n\right)-\left(u_{n+1}-u_n\right)\left(a_n-a_n^2-a_{n,x_1}\right)\\
&+a_na_{n+1}\left(u_{n+2}-u_{n+1}\right)+a_n\left(u_{n+1}-u_n\right)^2.
\end{aligned}
\end{equation}
To find a symmetry in the $y_1$ direction, we will use the invariance of system \eqref{gVolterra} and the associated non-local variables with respect to the simultaneous replacement of variables of the form
\begin{equation}\label{involutions}
a\leftrightarrow -b,\quad n\leftrightarrow -n, \quad x_i\leftrightarrow y_i,\quad Q\leftrightarrow H.
\end{equation}
It is clear from this reasoning that the sought symmetry is given by:
\begin{align*}
a_{n,y_2}=&2a_n\left(b_{n+1}-b_n\right)Q_n+a_n\left(b_{n+1,y_1}+b_{y_1}\right)-a_n\left(b_{n+1}-b_n\right)^2\\
&-a_nb_{n+1}(u_{n+1}-u_n)-a_n(u_{n+1}-u_n)-a_nb_n(u_n-u_{n-1}),\\
b_{n,y_2}=&b_{n,y_1y_1}-2b_nb_{n,y_1}-2\left(b_nQ_n\right)_{y_1}-b_n\left(u_{n,y_1}-u_{n-1,y_1}\right)+b_n\left(u_n-u_{n-1}\right)^2\\
&+b_nb_{n-1}\left(u_{n-1}-u_{n-2}\right)+\left(u_n-u_{n-1}\right)\left(2b_nQ_n+b_n^2-2b_{n,y_1}+b_n\right),\\
u_{n,y_2}=&u_{n,y}-2b_nQ_n(u_n-u_{n-1})-(u_n-u_{n-1})\left(b_n^2-b_{n,y_1}+b_n\right)\\
&-b_n\left(u_n-u_{n-1}\right)^2-b_nb_{n-1}\left(u_{n-1}-u_{n-2}\right).
\end{align*}
It is easily obtained from \eqref{symmfor3}.

Using reduction \eqref{reduction} we obtain the symmetries of lattice \eqref{newlattice} in the direction of $x_1$ from the found symmetries of system \eqref{gVolterra}:
\begin{align}\label{shortx1}
u_{n,x_2}=u_{n,x_1x_1}-2u_nu_{n,x_1}+u_n^2+1-2(u_n^2-u_{n,x_1}+1)\bar{H}_n
\end{align}
and in the direction of $y_1$:
\begin{align}\label{shorty1}
u_{n,y_2}=u_{n,y_1y_1}-2u_nu_{n,y_1}+u_n^2+1-2(u_n^2-u_{n,y_1}+1)\bar{Q}_n,
\end{align}
where $\bar{H}_n=(T-1)^{-1}D_{x_1}\log \frac{u_{n,x_1}-u^2_n-1}{u_{n+1}-u_n}$, $\bar{Q}_n=(T-1)^{-1}D_{y_1}\log \frac{u_{n+1,y_1}-u^2_{n+1}-1}{u_{n+1}-u_{n}}$.

Note that the symmetries \eqref{shortx1} and \eqref{shorty1} depend significantly on the discrete parameter $n$, since they contain variables with shifted arguments.
Now our aim is to rewrite them as coupled systems with two unknouns similar to \eqref{csVolterra}. Let us begin with \eqref{shortx1}. Firstly we concentrate on the shifted equation of the form \eqref{shortx1}:
\begin{align*}
u_{n-1,x_2}=u_{n-1,x_1x_1}-2u_{n-1}u_{n-1,x_1}+u_{n-1}^2+1-2(u_{n-1}^2-u_{n-1,x_1}+1)\bar{H}_{n-1}
\end{align*} 
One can replace the nonlocality $\bar{H}_{n-1}$ due to the relation 
\begin{align*}
\bar{H}_{n-1}=\bar{H}_{n} - D_x\log \frac{u_{n-1,x_1}-u_{n-1}^2-1}{u_{n}-u_{n-1}}. 
\end{align*} 
Afterwards the shifted equation takes the form
\begin{equation}\label{n-1}
\begin{aligned}
u_{n-1,x_2}=&-u_{n-1,x_1x_1}  +2(u_{n-1,x_1}-u_{n-1}^2-1)\bar{H}_{n}-\frac{2u_{n-1,x_1}^2}{u_n-u_{n-1}}\\
&+
\frac{ 2(u_{n-1,x_1}-u_{n-1}^2-1)u_{n,x_1}}  {u_n-u_{n-1}}+
\frac{2(u_{n}u_{n-1}+1)u_{n-1,x_1}}{u_n-u_{n-1}}      +u_{n-1}^2+1,
\end{aligned} 
\end{equation}
where the nonlocality $\bar H_n$ is given by
\begin{align*}
D_{y_1}\bar{H}_{n}=- D_{x_1} \frac{u_{y_1}-uu_{-1}-1}{u_{n}-u_{n-1}}. 
\end{align*} 
Thus finally we arrive at a coupled system for $u:=u_n$ and $v:=u_{n-1}$:
\begin{equation}\label{csx7}
\begin{aligned}
u_{x_2}=&u_{x_1x_1}-2uu_{x_1}+u^2+1-2(u^2-u_{x_1}+1)\bar{H}\\
v_{x_2}=&-v_{x_1x_1}  +2(v_{x_1}-v^2-1)\bar{H}-\frac{2v_{x_1}^2}{u-v}\\
&+
\frac{ 2(v_{x_1}-v^2-1)u_{x_1}}{u-v}+
\frac{2(uv+1)v_{x_1}}{u-v}      +v^2+1,
\end{aligned} 
\end{equation}
where $D_{y_1}\bar{H}=- D_{x_1} \frac{u_{y_1}-uv-1}{u-v}.$ Obviously system \eqref{csx7} does not contain any variable with shifted values of $n$.

The second order symmetry of the lattice \eqref{newlattice}  in another direction can also be transformed into a coupled system. For this we first exclude the variable $\bar Q_n$ according to the formula
\begin{align*}
\bar{Q}_{n}=\bar{Q}_{n-1} - D_y\log \frac{u_{ny_1}-u_{n}^2-1}{u_{n}-u_{n-1}}. 
\end{align*}
and rewrite \eqref{shorty1} as follows
\begin{align*}
u_{ny_2}=& -u_{ny_1y_1}+2(u_{ny_1}-u_n^2-1)\bar{Q}_{n-1}+\frac{2u^2_{n,y_1}}{u_n-u_{n-1}}-\\
&\frac{2(u_{ny_1}-u_n^2-1)u_{n-1,y_1}}{u_n-u_{n-1}}-\frac{2(u_nu_{n-1}+1)u_{ny_1}}{u_n-u_{n-1}}+u_n^2+1.
\end{align*}
The nonlocality satisfies the equation
\begin{align*}
D_x\bar{Q}_{n-1}=D_y\log \frac{u_{n-1,x}-u_{n}u_{n-1}-1}{u_{n}-u_{n-1}}. 
\end{align*}
Now we are ready to write down the desired coupled system for the functions $u:=u_n$, $v:=u_{n-1}$
\begin{equation}\label{csy8}
\begin{aligned}
u_{y_2}=& -u_{y_1y_1}+2(u_{y_1}-u^2-1)\bar{Q}+\frac{2u^2_{y_1}}{u-v}\\
&-\frac{2(u_{y_1}-u^2-1)v_{y_1}}{u-v}-\frac{2(uv+1)u_{y_1}}{u-v}+u^2+1.\\
v_{y_2}=&v_{y_1y_1}-2vv_{y_1}+v^2+1+2(v_{y_1}-v^2-1)\bar{Q},\\
\end{aligned} 
\end{equation}
where $D_x\bar Q=D_y\left(\frac{v_x-uv-1}{u-v}\right)$.

\bigskip
\noindent
\textbf{3.2.} In this section we construct the third order symmetries of system \eqref{gVolterra}. To this end we use the following two systems of equations:
\begin{equation}\label{ZakharovShabatx3}
\begin{aligned}
&\partial_{x_3}B_1-\partial_{x_1}B_3+[B_1,B_3]=0,\\
&\partial_{x_3}C_1-\partial_{y_1}B_3+[C_1,B_3]=0
\end{aligned}
\end{equation}
and
\begin{equation}\label{ZakharovShabaty3}
\begin{aligned}
&\partial_{y_3}C_1-\partial_{y_1}C_1+[C_1,C_3]=0,\\
&\partial_{y_3}B_1-\partial_{x_1}C_1+[B_1,C_3]=0.
\end{aligned}
\end{equation}

Here operators $B_3$ and $C_3$ are found by virtue of formulas \eqref{+-} and \eqref{Lk+Mk-}. For example, operator $B_3$ has the form:
\begin{equation}\label{B3}
\begin{aligned}
B_3=&-a_na_{n+1}a_{n+2}T^3+a_na_{n+1}(H_n+H_{n+1}+H_{n+2})T^2\\
&-a_n\left(a_{n+1}V_{n+2}+a_{n}V_{n+1}+a_{n-1}V_{n}+H_{n+1}^2+H_n^2+H_{n}H_{n+1}\right)T\\
&+a_na_{n+1}a_{n+2}-a_na_{n+1}(H_n+H_{n+1}+H_{n+2})\\
&+a_n\left(a_{n+1}V_{n+2}+a_{n}V_{n+1}+a_{n-1}V_{n}+H_{n+1}^2+H_n^2+H_{n}H_{n+1}\right).
\end{aligned}
\end{equation}
Similar to the previous case, we substitute explicit expressions of the operators $B_1$, $C_1$ and $B_3$ into system \eqref{ZakharovShabatx3} and obtain an overdetermined system of equations from which we find the symmetry of system \eqref{gVolterra} in the direction of $x_1$:
\begin{align*}
a_{n,x_3}=&-a_{n,x_1x_1x_1}-a_na_{n+1}u_{n+2,x_1}+2a_nu_{n+1,x_1x_1}+a_nu_{n,x_1x_1}+a_n\left(2a_n+3H_n\right)u_{n,x_1}\\
&+3D^2_{x_1}\left(a_nH_n\right)-D_{x_1}\left(a_n^3\right)+3D_{x_1}\left(u_{n+1}-u_n\right)\left(2a_nH_n-a_{n,x_1}\right)\\
&+3D_{x_1}\left(a_n^2H_n-a_na_{n-1}V_n-a_nH^2_n-a_nu_{n+1,x_1}-a_na_{n,x_1}\right)-a_n\left(u_{n+1}-u_n\right)^3\\
&+\left(3a_nH_n-2a_n^2-3a_{n,x_1}\right)\left(u_{n+1}-u_n\right)^2-3a_na_{n+1}\left(u_{n+2}-u_{n+1}\right)\left(u_{n+1}-u_n\right)\\
&-3a_n\left(u_{n+1,x_1}-u_{n,x_1}\right)-a_n\left(a_n^2-3a_nH_n+3H_n^2+3a_{n-1}V_n+7a_{n,x_1}\right)\left(u_{n+1}-u_n\right)\\
&-a_na_{n+1}\left(u_{n+2}-u_{n+1}\right)^2+a_na_{n+1}\left(a_{n+1}+a_{n+2}+3H_n-2a_n\right)\left(u_{n+2}-u_{n+1}\right)\\
&-\left(2a_na_{n+1,x_1}+3a_{n+1}a_{n,x_1}\right)\left(u_{n+2}-u_{n+1}\right)-a_na_{n+1}a_{n+2}\left(u_{n+3}-u_{n+1}\right)\\
&+\left(3a_{n,x_1}-3a_nH_n-2a_n^2+a_na_{n+1}\right)u_{n+1,x_1},\\
b_{n,x_3}=&b_n\left[a_{n,x_1x_1}-a_{n-1,x_1x_1}-3D_{x_1}(a_nH_n)+D_{x_1}\left(a_n\left(u_{n+1}-u_n\right)+a_{n-1}\left(u_{n-1}-u_n\right)\right)\right.\\
&\left.-a_{n-1}\left(u_{n-1}-u_n\right)^2+a_n\left(u_{n+1}-u_n\right)^2-3H_n\left(a_n-a_{n-1}\right)+a_na_{n+1}\left(u_{n+2}-u_{n+1}\right)\right.\\
&\left.+\left(u_{n+1}-u_n\right)\left(a_{n,x_1}+2a_n^2-a_na_{n-1}-3a_nH_n\right)+a_n^3-a_{n-1}^3+3a_na_{n,x_1}\right.\\
&\left.+\left(u_{n-1}-u_n\right)\left(a_{n-1,x_1}+2a_{n-1}^2-3a_na_{n-1}+3a_{n-1}H_n\right)-3a_{n-1,x_1}H_n\right.\\
&\left.+3\left(a_n-a_{n-1}\right)\left(a_{n-1,x_1}-a_na_{n-1}-H_n^2-a_{n-1}V_n\right)\right],\\
u_{n,x_3}=&\left(u_{n+1}-u_n\right)\left(3D_{x_1}\left(a_nH_n\right)-a_{n,x_1x_1}-a_n^3\right)-\left(u_{n+1}-u_n\right)^2\left(2a_n^2-3a_nH_n+2a_{n,x_1}\right)\\
&-3a_n\left(u_{n+1}-u_n\right)\left(H_n^2-a_nH_n+a_{n-1}V_n+a_{n,x_1}\right)-a_na_{n+1}a_{n+2}\left(u_{n+3}-u_{n+2}\right)\\
&-a_n\left(u_{n+1}-u_n\right)\left(u_{n+1,x_1}-u_{n,x_1}\right)-\left(u_{n+2}-u_{n+1}\right)\left(a_na_{n+1,x_1}+2a_{n+1}a_{n,x_1}\right)\\
&-a_n\left(u_{n+1}-u_n\right)^3-a_na_{n+1}\left(u_{n+2}-u_{n+1}\right)\left(2a_n-a_{n+1}-3u_n-3H_n+u_{n+2}\right).
\end{align*}
Due to the invariance of system \eqref{gVolterra} and the nonlocal variables associated with it, according to the following replacement 
\begin{equation}\label{involutions-x3}
a\leftrightarrow - b,\quad n\leftrightarrow -n, \quad x_i\leftrightarrow y_i,\quad Q\leftrightarrow H,\quad U\leftrightarrow V.
\end{equation}
we can easily obtain the symmetry of system \eqref{gVolterra} in the $y_1$ direction. We omit these computations.

Finally, by virtue of reduction \eqref{reduction}, we obtain two symmetries of lattice \eqref{newlattice} in the direction of $x_1$
\begin{align*}
u_{n,x_3}=&-u_{n,x_1x_1x_1}+3u_nu_{n,x_1x_1}+3u_{n,x_1}+\left(6u_nu_{n,x_1}-3u_{n,x_1x_1}\right)\bar{H}_n\\
&-3(u_n^2-u_{n,x_1}+1)\left(\bar{V}_n-\bar{H}_{n,x_1}-\bar{H}_n^2\right)-u_n^2-1
\end{align*}
and correspondingly in the direction of $y_1$
\begin{align*}
u_{n,y_3}=&-u_{n,y_1y_1y_1}+3u_nu_{n,y_1y_1}+3u_{n,y_1}+\left(6u_nu_{n,y_1}-3u_{n,y_1y_1}\right)\bar{Q}_n\\
&-3(u_n^2-u_{n,y_1}+1)\left(\bar{U}_n-\bar{Q}_{n,y_1}-\bar{Q}_n^2\right)-u_n^2-1,
\end{align*}
where $\bar{V}_n=(T-1)^{-1}D_{x_1} \left(\frac{u_{n,x_1}-u^2_n-1}{u_{n+1}-u_n}-u_n-\bar{H}_n\right),$ $\bar{H}_n=(T-1)^{-1}D_{x_1}\log \frac{u_{n,x_1}-u^2_n-1}{u_{n+1}-u_n}$, \linebreak $\bar{U}_n=(1-T)^{-1}D_{y_1} \left(\frac{u_{n+1,y_1}-u^2_{n+1}-1}{u_{n+1}-u_{n}}+u_{n+1}+\bar{Q}_{n+1}\right),$  $\bar{Q}_n=(T-1)^{-1}D_{y_1}\log \frac{u_{n+1,y_1}-u^2_{n+1}-1}{u_{n+1}-u_{n}}$.

\section{Construction of exact solutions to the coupled systems via integrable reductions of the dressing chains}

Now we discuss how the dressing chain can be used to construct explicit solutions to the coupled systems. As an illustrative example we take the system 
\begin{equation}\label{csVolterraR}
\begin{aligned}
&u_{t}=u_{xx}+\left(u^2+2uV\right)_x,\\
&v_{t}=-v_{xx}+\left(V^2\right)_y+(2uv)_x, \quad V_y=v_x,
\end{aligned}
\end{equation}
corresponding to the Volterra lattice (here $u:=u_n$ and $v:=v_n$)
\begin{equation*}%\label{Volterra}
u_{ny}=u_n(v_n-v_{n+1}), \qquad v_{nx}=v_n(u_{n-1}-u_n).
\end{equation*}

\begin{equation*}
u_{ny}=u_n(v_n-v_{n+1}), \qquad v_{nx}=v_n(u_{n-1}-u_n).
\end{equation*}
Let us consider its reduction
\begin{equation}\label{reduction2}
u_{y}=-uv, \qquad   v_{x}=uv      
\end{equation}
obtained due to cutting-off constraint $u_{-1}=0$, $v_{1}=0$. Here the sought functions are $u:=u_0$ and $v:=v_0$. The functions $I=\frac{u_x}{u}-u$ and $J=\frac{v_y}{v}+v$ are characteristic integrals of the system. Indeed, it is checked straightforwardly that the necessary conditions of the integrals $D_yI=0$ and $D_xJ=0$ hold. Therefore we have a system of differential equations (Bernoulli equations)
\begin{equation}\label{ODE}
\frac{u_x}{u}-u=f_1(x), \qquad   \frac{v_y}{v}+v=f_2(y)      
\end{equation} 
for searching solution to system \eqref{reduction2}, where $f_1$ and $f_2$ are arbitrary functions.

It is easy to check that general solution of the system can be parametrized in the following form 
\begin{equation}\label{solution}
u(x,y)=-\frac{W_x}{W}=\frac{\rho'(x)}{\varphi(y)-\rho(x)}, \quad  v(x,y)=\frac{W_y}{W}=\frac{-\varphi'(y)}{\varphi(y)-\rho(x)}      
\end{equation}
where $W=\varphi(y)-\rho(x)$ and $W_x$, $W_y$ denote the derivatives of $W$ with respect to $x$ and $y$. Here the functions $\varphi(y)$, $\rho(x)$ are chosen arbitrarily. Note that $W_{xy}=0.$

Now we assume that functions $u(x,y)$ and $v(x,y)$ depend on one more independent variable $t$ due to system \eqref{csVolterraR}. In other words we have $\varphi=\varphi(y,t)$ and $\rho=\rho(x,t)$. Then by integrating equation $V_y=v_x$ we derive an explicit expression for the nonlocality
\begin{equation}\label{nonlokV}
V=-\int \frac{\varphi_y\rho_xdy}{W^2}=\frac{\rho_x}{W}+R(x,t)=-u+R(x,t).
\end{equation}
Let us set $R(x,t)=0$ for simplicity. Then we get
\begin{equation}\label{nonlocalityV}
V=-u.
\end{equation}
Afterward the coupled system \eqref{csVolterra} turns into 
\begin{equation}\label{csFinal}
u_t=u_{xx}-2uu_x, \qquad v_t=-v_{xx}+2uu_y+2(uv)_x.
\end{equation}
 Using the substitution \eqref{solution}, we reduce the system \eqref{csFinal} to an overdetermined system of equations with a single sought function $W$.
To apply the substitution \eqref{solution} we have to use the explicit representations of the derivatives of $u$ and $v$:
\begin{equation}\label{explicitderivatives}
\begin{aligned}
&u_y=-v_x=\frac{W_x W_y}{W^2},\quad u_x=-\frac{W_{xx}}{W}+\frac{W_x W_y}{W^2},  \\
&u_{xx}=-\frac{W_{xxx}}{W}+3\frac{W_x W_{xx}}{W^2}-2\frac{W_x^3}{W^3}, \\
&u_t=-\frac{W_{xt}}{W}+\frac{W_x W_t}{W^2}, \quad  v_t=\frac{W_{yt}}{W}-\frac{W_y W_t}{W^2},\\
&v_{xx}=-\frac{W_{xx}W_y}{W^2}+2\frac{W_x^2W_y}{W^3}.
\end{aligned}
\end{equation}

The formulas above allow us to bring system \eqref{csFinal} to the following form
\begin{equation}\label{Usystem}
\begin{aligned}
&WW_{ty}=W_yW_{t}-W_{xx}W_y,  \\
&WW_{tx}=WW_{xxx}+W_xW_{t}-W_xW_{xx}.
\end{aligned}
\end{equation}
The next step is to solve it explicitly. Let us start with the second equation of \eqref{Usystem}. First, we represent the equation as
\begin{equation*}
\frac{W W_{xt}-W_x W_t}{W^2}=\frac{WW_{xxx}-W_xW_{xx}}{W^2}.
\end{equation*}
Then integrating the latter we get
\begin{equation*}
\frac{W_t}{W}=\frac{W_{xx}}{W}+g(t,y).
\end{equation*}

Now we simplify the first equation in \eqref{Usystem}     due to the relation 
\begin{equation}\label{d2}
W_t=W_{xx}+g(t,y)W
\end{equation}
and obtain an equation of the form
\begin{equation}\label{d3}
W_{yt}=g(t,y)W_y.
\end{equation}
If we apply the operator $D_y$ of the total differentiation with respect to $y$ to both sides of \eqref{d2} and then simplify it in virtue of equation $W_{xxy}=0$ we arrive at the relation
\begin{equation}\label{d4}
W_{yt}=g(t,y)W_y+g_y(t,y)W.
\end{equation}
Comparing relations \eqref{d3} and \eqref{d4} gives $g_y(t,y)=0$ or, the same,
\begin{equation}\label{d5}
g(t,y)=g(t).
\end{equation}

Analyzing the reasoning above we can conclude that the desired function $W=W(x,y,t)$ is a solution to the system
\begin{equation}\label{finalsyst}
\begin{aligned}
&W_{xy}=0,  \\
&W_{t}=W_{xx}+ g(t)W, \\
&W_{ty}=g(t)W_y.
\end{aligned}
\end{equation}
Obviously the third equation of the system is easily integrated, since it is of the form $\frac{\partial}{\partial t}\ln W_y=g(t)$. Hence it implies $\ln W_y= \ln G(t) +\ln F_1(y)$, where $\ln G(t)=\int_0^t g(\tau)d\tau$ and the constant of integration $\ln F_1(y)$ does not depend on $x$ due to the first equation in \eqref{finalsyst}. Then we integrate the obtained equation $W_y=G(t)F_1(y)$ with respect to $y$. It is convenient to present the result in the form   
\begin{equation}\label{d6}
W=G(t)(F(y)+S(x,t)),
\end{equation}
where $F(y)=\int_0^y F_1(z)dz$.

No we substitute \eqref{d6} into the second equation of \eqref{finalsyst}. After a slight simplification we obtain the heat equation for $S(x,t)$
\begin{equation}\label{d7}
S_t=S_{xx}.
\end{equation}
Therefore general solution to the system \eqref{finalsyst} is given by \eqref{d6} with arbitrary $G(t)$ and $F(y)$ and with an arbitrary solution $S(x,t)$ to \eqref{d7}. In other words we have 

{\bf Theorem 2}. {\it Assume that $S(x,t)$ is a solution to the equation \eqref{d7} and $F(y)$ is an arbitrary smooth function, then functions defined due to the rule 
\begin{equation}\label{d8}
\begin{aligned}
&u(x,y,t)=-\frac{\partial}{\partial x} \ln (S(x,t)+F(y)),\\
&v(x,y,t)=\frac{\partial}{\partial y} \ln (S(x,t)+F(y)),\\
&V(x,y,t)=\frac{\partial}{\partial x} \ln (S(x,t)+F(y)),
\end{aligned}
\end{equation}
give a solution to the coupled system \eqref{csVolterra}}

The statement of the theorem 1 is easily verified by a simple substitution.

As is known, the solution of the heat conduction equation \eqref{d7} is given in a closed form according to the Poisson formula 
\begin{equation}\label{Poisson}
S(x,t)=\frac{1}{2\sqrt{\pi t}}\int_{R}S_0(\xi)e^{-\frac{(x-\xi)^2}{4t}}d\xi,
\end{equation}
where $\left.S\right|_{t=0}=S_0(x)$ is a continuous and bounded function. Therefore solution \eqref{d8} of coupled system \eqref{csVolterra} depends on two arbitrary functions $S_0(x)$ and $F(y)$.

\section{The second example}
Taking a linear combination of two symmetries, we find a coupled system that depends symmetrically on $x$ and $y$
\begin{equation}\label{csVolterra30}
\begin{aligned}
&u_{n,s}=\lambda u_{xx}+ \mu u_{yy}+\lambda\left(u^2+2uV\right)_x    +\mu\left( U^2\right)_x +\mu(2uv)_y,\\
&\qquad V_y=v_x,\quad  \lambda\neq0\\
&v_{n,s}=\ -\lambda v_{xx} -\mu v_{yy}+\lambda \left(V^2\right)_y+\lambda(2uv)_x +\mu\left(v^2+2v U\right)_y,\\
& \qquad  U_x=u_y, \quad\mu\neq0.
\end{aligned}
\end{equation}
The boundary conditions $u_{-1}=0$ and $v_{1}=0$ imposed on the Volterra chain are compatible with all symmetries.
Therefore to construct solutions of system \eqref{csVolterra3} one can use the same ansatz
\begin{equation}\label{solution2}
u(x,y)=-\frac{W_x}{W}, \qquad  v(x,y)=\frac{W_y}{W}      
\end{equation}
as in the previous example. 
Here $W=\varphi(y)-\rho(x)$ with arbitrary functions $\varphi(y)$, $\rho(x)$ and $W_x$, $W_y$ denote the derivatives of $W$ with respect to $x$ and $y$. . We choose the nonlocalities as $V=-u$ and $U=-v$.

As a result, we arrive at the following system of equations:
\begin{equation}\nonumber
\begin{aligned}
W_{sx}-&\lambda W_{xxx}- \mu W_{xyy}-\frac{W_xW_s}{W}      +\lambda\frac{W_{xx}W_x}{W}    +2\mu\frac{W_{xy}W_y}{W}-\\
&\mu\frac{W_{x}W_{yy}}{W}=0,\\
W_{sy}+&\lambda W_{xxy}+ \mu W_{yyy}-\frac{W_yW_s}{W}      -\mu\frac{W_{yy}W_y}{W}    -2\lambda\frac{W_{xy}W_x}{W}+\\
&\lambda\frac{W_{y}W_{xx}}{W}=0.
\end{aligned}
\end{equation}
We integrate the first equation with respect to $x$ and the second one integrate with respect to $y$.  Finally we get one and the same equation
\begin{equation}\label{csVolterra5}
W_{s}-\lambda W_{xx}+ \mu W_{yy}-g(s)W=0.    
\end{equation}
It is simplified by the linear transformation $W=h(s)H$, where $h(s)$ is a solution to the equation $h'=gh$. The transformation brings it to a simple equation
\begin{equation}\label{heat}
H_{s}=\lambda H_{xx}- \mu H_{yy}.    
\end{equation}

{\bf Theorem 3}. {\it Assume that $H(x,y,s)$ is an arbitrary solution to the equation \eqref{heat}. Then the functions given in the following way
\begin{equation}\label{exactsolution}
\begin{aligned}
&u=-\frac{\partial}{\partial x}\ln (H),  \quad
v=\frac{\partial}{\partial y}\ln (H), \\
&U=\frac{\partial}{\partial y}\ln (H),\quad
V=-\frac{\partial}{\partial x}\ln (H)
\end{aligned}
\end{equation}
define a solution to the coupled system \eqref{csVolterra30}.}

Recall that the solution of the heat equation \eqref{heat} with $\lambda=-\mu=1$ is given in a closed form according to the Poisson formula 
\begin{equation*}
H(x,y,s)=\frac{1}{4\pi s}\int_{-\infty}^{\infty}\int_{-\infty}^{\infty} H_0(\xi,\eta)e^{-\frac{(x-\xi)^2+(y-\eta)^2}{4s}}d\xi d\eta,
\end{equation*}
where $\left.H\right|_{s=0}=H_0(x,y)$ is an arbitrary continuous and bounded function. Therefore, formulas \eqref{exactsolution} define a solution to coupled system \eqref{csVolterra30} depending on an arbitrary function of two variables $H_0(x,y)$.

\section{Construction of a particular solution of lattice (1.8).} 

In this secton we construct a particular solution of lattice \eqref{newlattice}, in which we set $u_{n+1}=i$, $u_{n-1}=-i$, $u:=u_n$:
\begin{equation}\label{lattice0}
u_{xy}=\frac{2uu_{x}u_{y}}{u^2+1}.
\end{equation}
Here we consider $n$ as an arbitrary but fixed integer. Next we will use the integrals of the lattice \eqref{newlattice} (see \cite{HabSakieva24}), which for $u_{n+1}=i$, $u_{n-1}=-i$ have the form:
\begin{equation}\label{Integrals}
J=\frac{u_{x}}{u^2+1}, \qquad I=\frac{u_{y}}{u^2+1}.
\end{equation}
Recall that the function $J(u,u_{x},u_{xx},u_{xxx},\ldots)$ is called a $y$-integral if the condition 
$$D_yJ(u,u_{x},u_{xx},u_{xxx},\ldots)=0$$ 
is satisfied. The $x$-integral is defined similarly. 

From the condition $D_yJ(u,u_{x},u_{xx},u_{xxx},\ldots)=0$ we find: 
$$J(u,u_{x},u_{xx},u_{xxx},\ldots)=f(x)$$
or in our case we have:
$$ \frac{u_{x}}{u^2+1}=f(x). $$
Integrating the last expression, we obtain:
$$u=\tan (F(x)+G(y)),$$
where $F'(x)=f(x)$, $G(y)$ are arbitrary functions.

Let's rewrite the found solution in general form
\begin{equation}\label{solution0}
u=\tan W(x,y),
\end{equation}
where $W(x,y)=F(x)+G(y)$.

Now we substitute \eqref{solution0} into the symmetry \eqref{csx7}:
\begin{align*}
u_{n,x_2}=u_{n,x_1x_1}-2u_nu_{n,x_1}+u_n^2+1-2(u_n^2-u_{n,x_1}+1)\bar{H}_n.
\end{align*}
of equation \eqref{newlattice}. First, we simplify this symmetry due to the restrictions $u_{n+1}=i$, $u_{n-1}=-i$, namely, we find out what the nonlocality $\bar{H}$ will be equal to in this case. Let's put $u:=u_n$, $x:=x_1$, $t:=x_2$, $y:=y_1$, $\bar{H}:=\bar{H}_n$, $u_{n+1}=i$, $u_{n-1}=-i$:
\begin{align}
& u_{t}=u_{xx}-2uu_{x}+u^2+1-2(u^2-u_{x}+1)\bar{H}, \label{symm0}\\
& D_{y}\bar{H}=- D_{x} \left(\frac{u_{y}+iu-1}{u+i}\right). \label{H0}
\end{align}
We integrate the last equality with respect to $y$ and find:
\begin{align*}
\bar{H}=-\frac{u_x}{u+i}-\varphi(x),
\end{align*}
where $\varphi(x)$ is arbitrary function. Here we will limit ourselves to considering case $\varphi(x)=0$. Due to what has been found, symmetry \eqref{symm0} can be written as:
\begin{align}\label{symm00}
u_t=u_{xx}-\frac{2u_x^2}{u+i}-2iu_x+u^2+1.
\end{align}
Let's substitute $u=\tan W(x,y,t)$ into \eqref{symm00}:
\begin{align*}
W_t=W_{xx}+i\left(2W_x^2-2W_x-i\right).
\end{align*}
Now we differentiate the last equation with respect to $x$:
\begin{align*}
W_{t,x}=W_{xxx}+i\left(4W_xW_{xx}-2W_{xx}\right).
\end{align*}
and make the substitution $W_x=\frac{1}{2}\left(-i\tilde{W}+1\right)$. Then our equation will be reduced to the Burgers equation:
\begin{align*}
\tilde{W}_t=\tilde{W}_{xx}+2\tilde{W}\tilde{W}_x.
\end{align*}
As is known, the Burgers equation, using the Cole-Hopf substitution $\tilde{W}=\alpha \frac{\bar{W}_x}{\bar{W}}$, is reduced to the heat equation:
\begin{align*}
\bar{W}_t=\bar{W}_{xx},
\end{align*}
when $\alpha=-1$. 
Then our desired particular solution will be written in the form
\begin{align*}
u&=\tan(W)=\tan\left(D_x^{-1}\left(\frac{1}{2}\left(-i\tilde{W}+1\right)\right)\right)=\tan\left(D_x^{-1}\left(\frac{1}{2}\left(i\frac{\bar{W}_x}{\bar{W}}+1\right)\right)\right)\\
&=\tan\left(\frac{i}{2}\ln(\bar{W}(x,y,t))+\frac{x}{2}+C(y,t)\right),
\end{align*}
where $\bar{W}(x,y,t)$ is a solution of the heat equation. 
Let us determine the dependence of function $\bar{W}(x,y,t)$ on variable $y$. To do this, we substitute the found solution into the $x$-integral $\frac{u_{y}}{u^2+1}=g(y,t)$, where $g(y,t)$ is some arbitrary function. We obtain:
\begin{align*}
\bar{W}(x,y,t)=\bar{F}(x,t)\bar{G}(y,t), 
\end{align*}
where $\bar{F}(x,t)$ is new arbitrary function, $\bar{G}(y,t)=e^{2iC(y,t)-2i\int g(y,t)dy}$.
Finally, the particular solution of lattice \eqref{newlattice} takes the form:
\begin{align*}
u=\tan\left(\frac{i}{2}\ln(\bar{F}(x,t)\bar{G}(y,t))+\frac{x}{2}+C(y,t)\right).
\end{align*}

Turning to the corresponding coupled system \eqref{csx7} we can convince that its solution is given by  
\begin{align*}
u&=\tan\left(\frac{i}{2}\ln(\bar{F}(x,t)\bar{G}(y,t))+\frac{x}{2}+C(y,t)\right),\\
v&=-i.
\end{align*}

\section*{Conclusion}

The problem of constructing explicit solutions for multidimensional integrable models is studied by many authors: Shabat, Zakharov, Novikov, Krichever, Manakov, Grinevich, Santini, Fokas, Taimanov, Konopelcnenko, Bogdanov, Ferapontov, Pavlov, Druma etc. A great variety of tools were suggested.   

Here we discussed the dressing chains method that provides an effective tool for constructing explicit solutions for integrable nonlinear PDE in the dimension 1+1 (see, for instance, \cite{ShabatYamilov91}, \cite{Smirnov22} and references therein). However in 3D some difficulties arise due to the  nonlocal variables (see \cite{LeznovShabatYamilov}). Examples considered in the article convince  that to overcome these difficulties one can use finite reductions of the dressing chans obtained by imposing cutting off constraints preserving integrability. Besides the degenerate boundary conditions related to reductions integrable in the sense of Darboux one can use also more general boundary conditions compatible with the integrability property of the lattices.

\end{document}